\documentclass[aps,prl,twocolumn]{revtex4-1}

\usepackage{amsmath}
\usepackage{graphicx,xcolor}
\usepackage{amsmath,amssymb,mathtools,slashed}
\usepackage{tikz-feynman}[luatex=true]
\usepackage{braket}
\usepackage{feynmp-auto}
\usepackage{simpler-wick}
\usepackage{kantlipsum}
\usepackage{float}
\usepackage{multirow}
\usepackage{rotating}

\usepackage[T1]{fontenc} 
\usepackage{etoolbox}
\usepackage{extarrows}
\usepackage{graphicx}
\usepackage{dcolumn}
\usepackage{bm}
\usepackage{wasysym}
\usepackage{verbatim}

\usepackage{empheq}
\usepackage{multirow}
 
\usepackage{balance}

\newcommand{\mpl}{M_\text{pl}}

\usepackage{subfigure}
\usepackage[colorlinks=true,urlcolor=blue,linkcolor=blue]{hyperref}
\usepackage[capitalise]{cleveref}
\usepackage{siunitx}
\usepackage{enumerate}
\newcommand{\be}{\begin{equation}}
\newcommand{\ee}{\end{equation}}
\newcommand{\bea}{\begin{eqnarray}}
\newcommand{\eea}{\end{eqnarray}}
\usepackage{array}
\usepackage{tcolorbox}
\usepackage{fontawesome}

\newcommand{\p}{\partial}

\newcommand{\bq}{\boldsymbol{q}}

\newcommand{\bB}{\boldsymbol{B}}

\newcommand{\bx}{\boldsymbol{x}}
\newcommand{\nint}[1]{\left\lfloor #1 \right\rceil}

\begin{document}


\title{QuGrav: Bringing gravitational waves to light with Qumodes }
\author{Dmitri E. Kharzeev} \affiliation{Center for Nuclear Theory, Department of Physics and Astronomy,
Stony Brook University, New York 11794-3800, USA}
\affiliation{Energy and Photon Sciences Directorate, Condensed Matter and Materials Sciences Division,
Brookhaven National Laboratory, Upton, New York 11973-5000, USA}
\author{Azadeh Maleknejad} \affiliation{Department of Physics,
Swansea University,
Singleton Park, Swansea,
SA2 8PP, UK}
\author{Saba Shalamberidze} \affiliation{Center for Nuclear Theory, Department of Physics and Astronomy,
Stony Brook University, New York 11794-3800, USA}

\begin{abstract}

We propose using qumodes, quantum bosonic modes, for detecting high-frequency gravitational waves via the inverse Gertsenshtein effect, where a gravitational wave resonantly converts into a single photon in a magnetized cavity. For an occupation number $n$ of the photon field in a qumode, the conversion probability is enhanced by a factor of $n+1$ due to Bose-Einstein statistics.  Unlocking this increased sensitivity  entails the ability to continuously prepare the qumode and perform non-demolition measurement on the qumode-qubit system within the qumode coherence time. Our results indicate that, at microwave frequencies and with existing technology, the proposed setup can attain sensitivities within 1.7 orders of magnitude of the cosmological bound. With anticipated near-future improvements, it has the potential to surpass this limit and pave the way for the first exploration of high-frequency cosmological gravitational wave backgrounds.  At optical frequencies, it can enhance the sensitivity of current detectors by one order of magnitude. That further enhances their potential in reaching the single-graviton level. 

\end{abstract}

\maketitle

{\bf{Introduction.}} 
At present, precise observations of gravitational waves (GW) are limited to frequencies below 10 kHz. The detection of higher-frequency GWs (HFGW) remains an open frontier, currently under active research and development.  These HFGWs may arise from cosmological or astrophysical sources; they may also be produced in a laboratory by conversion of high-intensity laser fields. Once detected, HFGWs can reveal a wealth of information about the early universe, particle physics, and the quantum nature of gravity. Remarkably, electromagnetic (EM) and GWs can interconvert in the presence of a static magnetic field—a phenomenon known as the Gertsenshtein effect \cite{Gertsenshtein}. Photon regeneration is a leading approach to probe HFGWs, which is based on this mechanism \cite{Ejlli:2019bqj}. In such experiments, GWs traverse a photon-shielded cavity permeated by a strong magnetic field, allowing for resonant graviton-to-photon conversion. 
The current active experiments are OSQAR \cite{OSQAR:2015qdv} and ALPS   \cite{Albrecht:2020ntd} at optical frequencies, and CAST  \cite{CAST:2017uph} at X-ray frequencies. In addition,  BabyIAXO and IAXO \cite{IAXO:2020wwp, IAXO:2019mpb} are under development, and JURA   \cite{Beacham:2019nyx} is proposed. In \cref{tab:experiments} we summarize these experiments. Some of these instruments operate well below the standard quantum limit. Ongoing research and development efforts aim to design next-generation instruments with enhanced power and sensitivity. That includes using stronger magnetic fields, bigger and higher quality cavities, and so on. For a recent review that covers these developments comprehensively, see \cite{Aggarwal:2025noe} and references therein.

In this letter, we propose a new approach to enhancing the sensitivity of cavity-based gravitational wave searches and reducing the signal-to-noise ratio.
Our approach is based on combining the following three ideas developed previously in different fields:  1) Photon (re)-generation, i.e. the resonant conversion of GWs into single photons in a background magnetic field; 2) Bosonic stimulation, i.e., an enhancement of a single boson production by a factor of $n+1$ when the bosonic state has an occupation number $n$; and 3) The use of qumodes (states with a fixed number of photons in cavities), which have been developed recently for bosonic quantum computing \cite{Blais:2020wjs,Stavenger:2022wzz,Araz:2024kkg}.

\begin{figure}     \includegraphics[height=3.5cm]{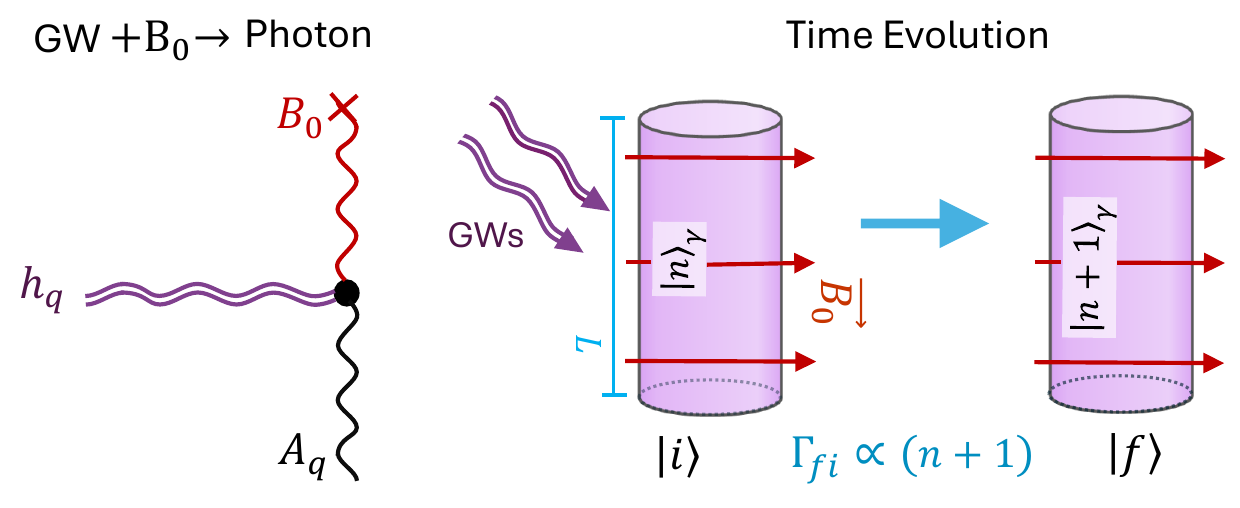} 
    \caption{Left: the inverse Gertsenshtein effect, and right: toy model of Qumode detector. 
    }
    \label{fig:illustration}
\end{figure}
  
The first of these ingredients underlies a number of current and planned experiments, as discussed above. The second one, \textit{bosonic stimulation}, has been established experimentally in the formation of Bose-Einstein condensates \cite{miesner1998bosonic}, and in light scattering in cold atom condensates \cite{lu2023bosonic}. The 3rd central pillar of our proposal is $n$-photon \textit{qumodes} \cite{qu-exp1,qu-exp2}; the measurement of photon number up to $\sim 100$ has already been demonstrated \cite{Cheng:2022wgy}, and is a key ingredient of bosonic quantum computation.

\begin{table}[H]
\centering
\small
\begin{tabular}{|l|l|c|c|c|c|c|c|c|}
\hline
 & \boldmath$f$ [Hz] & \boldmath$B_0$ [T] & \boldmath$L$ [m] & \boldmath$A$ [m$^2$] & \boldmath$\Gamma\!_{{\!D}}$ [mHz] & \boldmath$\varepsilon$ \\
\hline
OSQAR II       & $10^{15}$   & 9    & 14.3 & $5 \times 10^{-4}$  & $1.14$ & 0.9 \\
ALPS II     & $10^{15}$   & 5.3  & 106  & $10^{-3}$  & $10^{-3}$    & 0.75 \\
\textit{JURA}      & $10^{15}$   & 13   & 960  & $8 \times 10^{-3}$  & $10^{-6}$    & 1 \\
CAST       &  $10^{18}$ & 9    & 9.26 & $2.9 \times 10^{-3}$ & $0.15$ & 0.7 \\
IAXO                  & $10^{18}$  & 2.5  & 20   & 3.08                & $0.1$    & 1 \\
\hline
\textit{QuGrav}    & $10^{10}$    &  10 & $1 $    & 1 & $10^{-3}$ & 1                  \\
\hline 
\end{tabular}
\caption{Experimental parameters of the photon-regeneration setups. Here $\varepsilon$ is the single photon detection efficiency, and $\Gamma_D(f)$ is the dark count rate in a frequency bin of width $\Delta f$. 
}
\label{tab:experiments}
\end{table}

{\bf{Theoretical Setup.}}  The Maxwell-Einstein action describes the EM fields in curved space-time. Consider the space-time metric as
\begin{align}
g_{\mu\nu}=\eta_{\mu\nu}+h_{\mu\nu},
\end{align}
where  $\eta_{\mu\nu}={\rm diag}(-1,1,1,1)$ is the flat space-time metric, $\mu,\nu = 0,1,2, 3$, and $h_{\mu\nu}$ denotes small gravitational perturbations, i.e. $|h_{\mu\nu}|\ll 1$.  The EM-gravitational interaction Hamiltonian density at first order in $h_{\mu\nu}$ is 
\be\label{eq:L-int}
\mathcal{H}_{\rm int} = -\frac{1}{2} \eta^{\lambda\sigma}( h^{\mu\nu} - \frac12 h \eta^{\mu\nu}) F_{\mu\lambda}F_{\nu\sigma},
\ee
where $F_{\mu\nu}$ is the EM field  tensor. Throughout this work, we adopt natural units by setting $c=\hbar=\epsilon_0=1$.

Consider a cylindrical cavity of length $L$ and cross-sectional area $A$. We assume that the cavity is made of a perfect conductor and turn on a constant magnetic field in the $z$-direction inside it $
\bB(t,\bx) = B_0 \hat{z}$. 

At this point, we need to fix the gauge. The frame in which we measure the EM fields in the lab is the Detector Proper (DP) frame, i.e., the local inertial frame in which the detector is in free fall.  By contrast, theoretical descriptions often employ the Transverse-Traceless (TT) gauge, where test masses remain at fixed coordinate positions while the metric itself oscillates due to the passing wave. In this work, we adopt the TT gauge, i.e.
  $  h_{00} = h_{0i} = \p_i h_{ij} =0$,
where $i,j = 1,2,3$. We assume that $\bB_0$ is static in the TT frame. While, in general, the transformation between frames in the presence of GWs requires careful treatment, it is justified in the high-frequency regime $\omega L \gg v_s$ \cite{Ratzinger:2024spd}. Note that the sound velocity inside solids is typically small, i.e., $v_s <10^{-4}$.

\vspace{0.25cm}
{\bf{Photon-Graviton Transition Amplitude inside Cavity.}}   As the initial state, we consider a Qumode state made of an $n$-photon state at  a resonant frequency $\omega$
\begin{align}
   \ket {Q_{n} } =\ket{n(\omega)}_{\!\gamma},
   \label{eq:Qu-1}
\end{align}
where $\lvert n(\omega) \rangle_{\!_{\gamma}} =(b^{\dag}_{\omega})^{n}  \ket{0}_{\!\gamma}$.
This state is achievable with current experimental capabilities and is well-suited for detecting narrowband signals. To complement this, we also introduce a more involved qumode configuration to illustrate the theoretical potential and long-term prospects of this approach for broadband detection. The second state is a Qumode state made of an $n$-photon state at each resonant frequency of the cavity, $\omega_{l}=\frac{ \pi l}{L}$, i.e.
\begin{align}
   \overline{\ket{ Q_n }} =\prod_{l=1}^{\nint{2\Delta f L}} \otimes \ket{n(\omega_{l})}_{\!\gamma}.
    \label{eq:Qu-2}
\end{align}
where $\lvert n(\omega_l) \rangle_{\!_{\gamma}} =(b^{\dag}_{\omega_l})^{n}  \ket{0}_{\!\gamma}$, and  $\nint{2\Delta f L/c}$ is the number of resonant modes inside ($ f=\frac{\omega}{2\pi}$). The gravitational interaction \cref{eq:L-int} allows for the conversion between gravitons and photons. The transition amplitude from an initial state $\vert i \rangle$ to a different final state $\vert f \rangle$  is given by the S-matrix element 
\begin{align}
 \mathcal{A}_{fi} =\langle f \vert T\exp[-i \int d^4x \, \mathcal{H}_{\text{int}} ]\vert i \rangle,
 \label{eq:S-matrix}
\end{align}
where $T$ is the time ordering operator. Now we need to specify the GW signal interacting with the magnetic field in the cavity. In the following, we will consider a gravitational signal passing by, and here we consider three possibilities: i) a monochromatic GW signal, ii) a stochastic GW field background, and iii) a single-graviton. 
\newline

 \textbf{I) Incoming Coherent GWs: } Consider an unpolarized 
monochromatic GW with frequency $\omega_g$ as 
($\bq= \omega_g \hat{\bq}$)
\begin{align}
h_{ij}(t,\bx) = \sum_{\sigma=\pm}  \, e^{\sigma}_{ij}(\hat{\bq}) \,
 \hat{h}_{\omega_g} e^{iqx} + h.c.,
\end{align}
where $e^{\sigma}_{ij}(\hat{\bq})$ are the polarization tensors, $\sigma=\pm$ are the circular polarizations of the GWs, and $\hat{h}_{\omega_g}$ is a dimensionless complex-valued operator such that $\hat{h}_{\omega_g} \ket{\alpha}_{\!g} = h_{\omega}\ket{\alpha}_{\!g}$ where $\ket{\alpha}_{\!g}$ is a coherent GW state, $h_{\omega}$ is the mode function, and $\langle h_{ij}(t,\bx)h_{ij}(t,\bx) \rangle =2\lvert h_\omega \rvert^2$. 
The interaction Hamiltonian density can be written as
\begin{align}
\mathcal{H}_\text{int} = -B_0 \sin\theta_{\!\bq}  \sum_{\sigma}  \delta B_{i}(x) \, e^{\sigma}_{i}(\hat{\bq})  \,   \hat{h}_{\omega}  e^{iqx} ,
\end{align}
where $e^{\pm}_{j}(\hat{\bq}) = \frac{1}{\sqrt{2}}(\cos\theta_{\!\bq} \cos\phi_{\!\bq} \mp i \sin \phi_{\!\bq}, \cos\theta_{\!\bq} \sin\phi_{\!\bq} \pm i \cos\phi_{\!\bq}, -\sin\theta_{\!\bq})$ in cartesian coordinate. 

Here, the initial and final photon-graviton states are
\begin{align}
    \ket{i} = \ket{Q_n}_{\!\gamma} \ket{\alpha}_{\!g}, \quad \ket{f} = b^{\dag}_{\omega_g}\ket{Q_n}_{\!\gamma} \ket{\alpha}_{\!g}.
\end{align}
where a photon is created at frequency $\omega_g$, raising the occupation number at this frequency from $n$ to $n+1$. Given that $\lvert \bra{n+1} b^{\dag} \ket{n}\rvert^2 \propto (n+1)$, the transfer rate is  
\begin{align}
\Gamma_{fi}  =   2 (n+1)\,   Q  V   B^2_0   h^2_{\omega_g}\, \sin^2\theta_q,
\label{eq:rate-}
\end{align}
where $Q$ is the quality factor of the cavity (see \cref{fig:illustration}). Remarkably, due to Bose–Einstein statistics, once the frequency of the GW matches the frequency of the n-photon mode inside the cavity, the probability of graviton-to-photon conversion is enhanced by a factor of $n+1$. This process is the EM analogue to the stimulated absorption of gravitons in the massive quantum acoustic resonators, with the process being most effective at frequencies that match the resonator's natural frequency \cite{Tobar:2023ksi} 
Stimulated conversion in the context of dark matter searches has been demonstrated with $n=4$ photon states in a microwave cavity \cite{Agrawal:2023umy}. Our work opens a new direction by introducing stimulated graviton–photon conversion and systematically exploring different frequency ranges to identify the most promising regime for 
gravitational wave detection. While a bright coherent drive also enhances the generation rate by the factor 
$n+1$, it simultaneously introduces amplitude and phase noise. A quantitative comparison is given in the Supplemental Material.

 \textbf{II) Cosmic GW Background: }  A fraction of high-frequency gravitational waves is expected to originate from early-Universe phenomena such as inflation or phase transitions. Their total energy density is constrained by cosmological observations, particularly through limits on the effective number of relativistic species, $\Delta N_{\text{eff}}<0.3$, which gives  $h(f) < 2 \times 10^{-30} \, (\frac{1 \, \text{GHz}}{f}) \, \Delta N_{\text{eff}}^{\frac12} $.  If the GW background is dominated by a localized feature in frequency, it can be treated as a narrowband source, with the transition rate given by \cref{eq:rate-}. Otherwise, it generally exhibits a broadband spectrum, for which we compute the transition rate below.

 Consider an unpolarized 
stochastic background of GWs with cosmological origin. It can be described as 
\begin{align}
h_{ij}(x) = \sum_{\sigma=\pm} \int d\ln \omega \, e^{\sigma}_{ij}(\hat{\bq}) \, \int \frac{d^2\hat{\bq}}{2\pi} \,
 \hat{h}_{\omega} e^{iqx} + h.c.,
\end{align}
 where $\bq= \omega \hat{\bq}$, $h_{\omega}$ is the mode function, and $\langle h_{ij}(t,\bx)h_{ij}(t,\bx) \rangle =2\int d\ln\omega \lvert h_\omega \rvert^2$. 
Let $\Delta f$ denote the finite frequency range of the detector. For simplicity, assume the GW spectrum is flat across this interval. To improve the sensitivity in this case, we use \cref{eq:Qu-2} and have to sum over all the resonant modes. The total transfer rate is
\begin{align}
    \Gamma_{\text{tot}} 
     =   2 (n+1) \Delta f L V   B^2_0   h^2_{\omega}\, \sin^2\theta_q.
     \label{eq:Gamma-bb}
\end{align}


 \textbf{III) Single Gravitons: } Modern detector architectures can detect single photons, but building a detector sensitive to single gravitons remains an extraordinary challenge \cite{Dyson:2013hbl, Rothman:2006fp}. More precisely, if gravity is quantized, then the number of gravitons per de Broglie volume is 
\begin{align}
N_{\text{grav}} = n_{\text{grav}} \lambda_{\text{dB}}^3 = \frac{\pi \, h^2_{\omega}  \, \mpl^2}{2f^2},
\end{align}
where $\lambda_{\text{dB}}=f^{-1}$. As an example, a LIGO detection signal with $h_{\omega}\sim 10^{-25}$ at $f\sim 10^{2}$ Hz, includes $N_{\text{grav}} \sim 10^{29}$ gravitons. Requiring less than a graviton per de Broglie volume, we find
\begin{align}
 h_{\omega} \leq 2\times 10^{-31}   \, \Big(\frac{f}{10^{3} \, \text{GHz}}\Big),
\end{align}
where below that limit, the gravitational field is more accurately described as a highly dilute graviton gas rather than a classical gravitational wave. Notably, establishing the quantum nature of gravitational radiation demands measurements far more challenging than merely detecting individual gravitons \cite{Carney:2023nzz}.

Consider a single graviton propagating in direction $\hat{\bq}$ with frequency $\omega_g$, i.e. $\ket{1_{\omega_g}}_{\!g}$, as 
\begin{align}
h_{ij}(t,\bx) =  \sum_{\sigma=\pm} \, e^{\sigma}_{ij}(\hat{\bq}) \, h_{\omega_g}^{\sigma} \, \hat{a}_{\omega_g, \sigma} e^{-i\omega_g t+ i\bq.\bx} + h.c.,
\end{align}
where  $\hat{a}^{\dag}_{\omega,\sigma}$ and $\hat{a}_{\omega,\sigma}$ are the graviton creation and annihilation operators, $[\hat{a}_{\omega,\sigma},\hat{a}^{\dag}_{\omega',\sigma'}]=\delta_{\sigma\sigma'}\delta(\omega-\omega')$.
 Here we set the initial and final states as
 \begin{align}
     \ket{i} = \ket{Q_n}\ket{1_{\omega_g}}_{\!g}, \quad  \ket{f} = \hat{b}^{\dag}_{\omega_g}\ket{Q_n}\ket{0}_{\!g},
 \end{align}
and the transition rate is equal to \cref{eq:rate-}.

\vspace{0.25cm}
{\bf{Sensitivity and Detection Prospect.}} 
Here we compute upper limits
on the GW dimensionless characteristic strain, $h(f)$, based on the characteristics of each experiment, with the application of the
Qumode technology. These experiments detect the photon count rate $\Gamma$ within an energy band $\Delta f=(f_f-f_i)$, with a single photon detection efficiency $\varepsilon$, and over a cross-sectional area $A$. The background noise is quantified in terms of the dark count rate in that frequency bin, i.e., $ \Gamma_D(f)$. It is the rate at which the detector falsely registers a photon even when no real photon is present. See \cref{tab:experiments} for the experimental parameters of the photon-regeneration setups. In the following, we refer to the implementation of this technology in each GW detector by appending the prefix ``\textit{Qu}'' to its name. For example, the upgraded version of JURA incorporating this technology is denoted as \textit{Qu}JURA.  We call these updated experiments collectively as \textit{Qu}-detectors.

For completeness, we note that $n$-photon qumodes in the presence of a magnetic field also introduce a subleading noise contribution due to graviton emission, at a rate  $\Gamma_{\text{rev}}  =  \frac{ 2 n}{\mpl \sqrt{V \omega}}\,  Q V   B^2_0$. This process introduces a photon-number ambiguity, leading to an additional noise contribution of order $\frac{ 2 \Delta t}{\mpl \sqrt{V \omega}}\,  Q V   B^2_0$. As this effect is subleading compared to thermal and other dominant noise sources, we neglect it in the present analysis. 
While genuine photon-generation backgrounds, would also be subject to Bose enhancement, they are suppressed in the cryogenic, shielded regime considered here; the dominant background arises from detector read-out noise, which does not create photons in the mode and is therefore not Bose-enhanced (see the Supplemental Material).

   \textbf{Narrowband signal-to-noise ratio estimator (SNR):} For resonant detectors with narrow bandwidth sensitivity, a monochromatic gravitational field, or a GW background with a sharply peaked spectrum, SNR reduces to a contribution from a narrow frequency band 
 $\Delta f = \frac{f}{Q} = \frac{1}{L\mathcal{F}}$ where $\mathcal{F}$ is the finesse of the cavity. Now using \cref{eq:rate-} and demanding $\varepsilon \Gamma_{fi}=\Gamma_D$, we have
 \begin{align}
      h_{\text{noise}}(f) & \approx \frac{1.4 \times 10^{-27}}{\sqrt{n+1}} \left(\frac{1 \, \text{m}}{L}\right) \left(\frac{10 \, \text{T}}{ B_0}\right) \sqrt{\left(\frac{1 \, \text{m}^2}{A }\right) \, \left(\frac{10^5}{\mathcal{F}}\right)}
      \nonumber\\
      & \times \sqrt{\frac{1}{\varepsilon}\left(\frac{\Gamma_D}{1 \, \mu\text{Hz}}\right) \left(\frac{1\, \text{GHz}}{f}\right)}.
\label{eq:SNR-nb}
\end{align}

    \textbf{Broadband SNR: } Using the theoretical benchmark Qumode state \cref{eq:Qu-2} with \cref{eq:Gamma-bb} for detectors and signals with broad frequency ranges, we can quantify the SNR by demanding that $\varepsilon \Gamma_{\text{tot}}>\Gamma_D$; we find 

\begin{align}
 h_{\text{noise}}(f)  & \approx \frac{4.5 \times 10^{-25}}{\sqrt{(n+1)}}  \left(\frac{1 \text{m}}{L}\right) \left(\frac{10 \text{T}}{B_0 }\right) \sqrt{ \left(\frac{1 \, \text{GHz}}{\Delta f}\right) }   \nonumber\\
  & \times \sqrt{\frac{1}{\varepsilon}\left(\frac{\Gamma_D }{1 \, \mu\text{Hz}}\right)  \left(\frac{1 \, \text{m}^2}{A}\right)} .
  \label{eq:SNR-bb}
\end{align}

Here we can compare the Qumode implementation of a given detector with its standard single-photon detector 
\begin{align}
  h_{\text{noise}}^\text{single}/ h^{Qu}_{\text{noise}} = (n+1)^{-\frac12}.   
\end{align}

Consequently, the use of $n$-photon states reduces the noise-equivalent strain by a factor of $(n+1)^{-1/2}$.  For illustration, we benchmark our estimates with $n=100$, already demonstrated in the optical domain \cite{Cheng:2022wgy}, which improves the sensitivity by about an order of magnitude. While such large-$n$ states are still under development in the microwave regime, steady progress in cavity QED and superconducting systems suggests they represent a realistic long-term target. This highlights the potential of $n$-photon qumodes to significantly enhance sensitivity across a broad range of frequencies. The enhanced sensitivity, however, comes at the price of continuously rebuilding the qumode through its coupling to a single qubit, and performing a quantum non-demolition measurement on it within its coherence time \cite{Schuster:2007fwf}. Indeed, it has been demonstrated that an additional single photon can have a big effect on the qubit without ever being absorbed \cite{Schuster:2007fwf}. At present, the demonstrated qumode coherence time already exceeds 10 ms both for microwave \cite{Reagor:2013} and optical cavities \cite{Kessler}. However, at X-ray frequencies, the lifetime of 
$n$-photon states are short compared to the time needed for their preparation, making this technology currently unfeasible for CAST and IAXO.

\begin{figure}     \includegraphics[height=6cm]{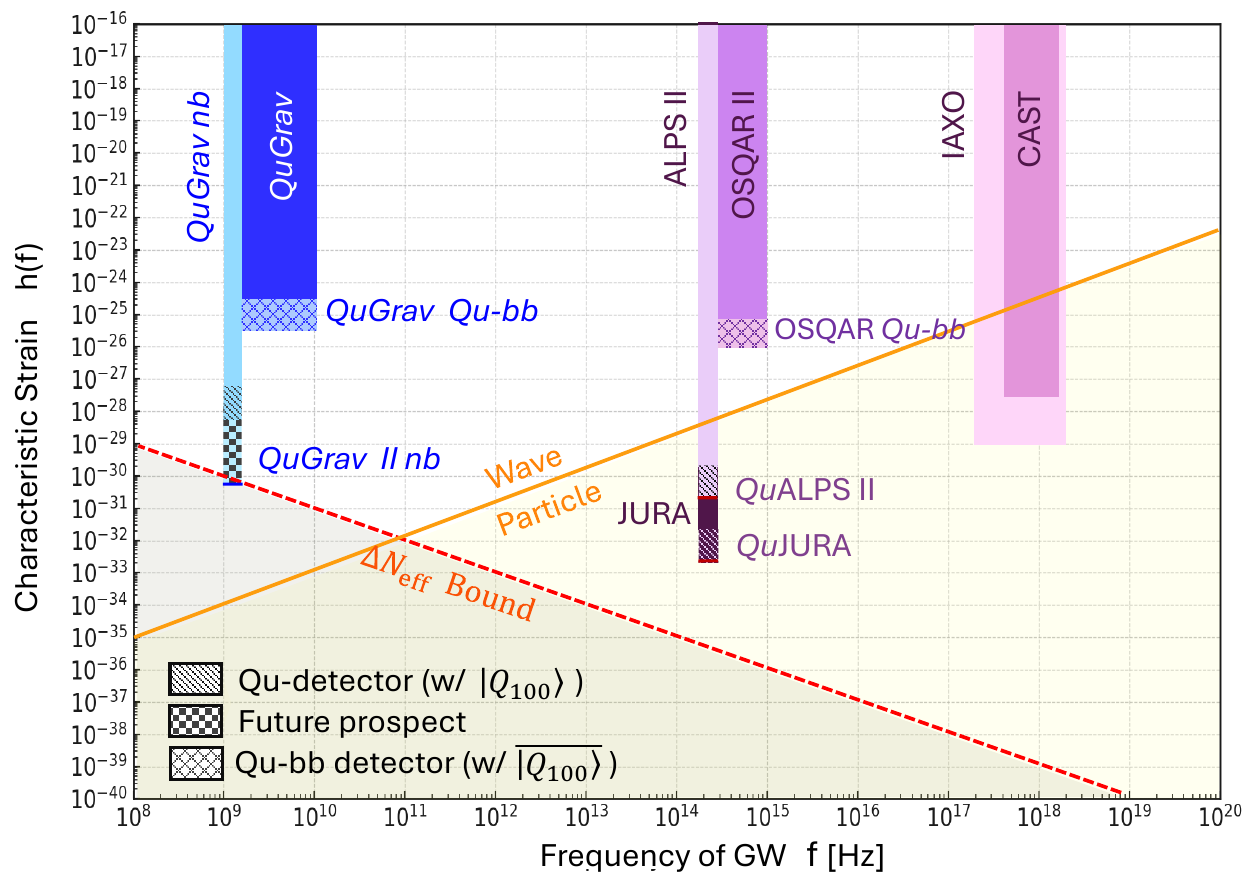} 
    \caption{Strain sensitivity of HFGW detectors based on photon regeneration. The area below the orange line corresponds to gravitational fields that behave like quantum particles (gravitons). The dashed gray curve denotes the $\Delta N_{\text{eff}}$ cosmological bound. Hatched regions indicate projected improvements in sensitivity from current experiments and proposals employing a 100-photon qumode ( $\ket{Q_{100}}$ in \cref{eq:Qu-1}) within the cavity.  The checkered (black-blue)  represents a sensitivity that relies on technological progress in cavity development anticipated in the near future, i.e., increasing S by 1.7 orders of magnitude (see \cref{eq:def-S}). The crossed regions indicate the enhanced sensitivity to broadband signals achievable with the futuristic broadband 100-photon Qumode $\overline{\ket{Q_{100}}}$ described in \cref{eq:Qu-2}.}
    \label{fig:detector}
\end{figure}
\textbf{$\bullet$ \textit{Qu}OSQAR II \& \textit{Qu}ALPS II} ($10^{6}$ GHz): The sensitivities of OSQAR II (active) and ALPS II  (in development) probing optical frequencies are respectively $h\approx 6 \times 10^{-26}$ and $h\approx 3 \times 10^{-30}$. One of the key advancements in ALPS II is the implementation of a Fabry-Perot (FP) cavity with  $\mathcal{F}=4\times 10^4$ to enhance the photon regeneration process. The proposed JURA experiment with  $\mathcal{F}=10^5$ aims to reach a strain sensitivity as low as $h\approx 2\times 10^{-32}$. \cite{Ejlli:2019bqj}. As we see in \cref{fig:detector}, the ALPS II and JURA are in the quantum particle limit. By employing $n=100$ Qumode states, the sensitivity can be enhanced by an order of magnitude, making the detectors even more responsive to signals of single gravitons.
\vskip 0.1cm

\textbf{$\bullet$ \textit{Qu}Grav} ($1-10$ GHz): Despite the technical challenges at GHz frequencies, rapid progress is underway, driven in part by dark photon and dark matter searches \cite{Lescanne:2020awk, Dixit:2020ymh}. The main challenge is that the photon's energy at this frequency is very low, i.e., around $10^{-6}$ eV (about $10$ mK). Therefore, to dominate over the thermal noise, we need cryogenic temperatures. In addition, even a micron-scale mechanical shift can detune the cavity. Therefore, it requires active feedback stabilization, temperature control, and vibration isolation, which makes it challenging. 
    Note, however, that the cavities coupled to superconducting qubits operate at mK temperatures, and there is a lot of ongoing work on the cryogenic microwave cavities motivated mainly by the needs of bosonic quantum computing, see e.g. \cite{qu-exp1,qu-exp2}.  The minimal frequency of a photon inside the cavity is $    f_{\text{min}} \sim   10^{-1} \,  \text{GHz} \, \big(\frac{L}{1 \, \, \text{m} }\big)  \, \big(\frac{1 \, \, \text{m}^2}{A}\big)$  \cite{Ringwald:2020ist}; hence, we will consider frequencies $f \gg f_{\text{min}}$.

    Here we make estimates for a toy detector operating with currently accessible parameters  $ B_0=10$ T, $\varepsilon=1$,  $\Gamma_D =1$ $\mu$Hz, $L=1 $ m, $\mathcal{F}=5 \times 10^5$, and $A=1$ $\text{m}^2$ operating at $mK$ temperature. Setting the resonant frequency at $f=10 f_{\text{min}} = 1-10$ GHz, we find the narrowband sensitivity to $h$ as
     \begin{equation}
     h_{ \text{noise}} \approx 6.3 \!\times \!10^{-29} S^{-1},
     \label{eq:h-micro}
     \end{equation}
where we define
\begin{align}
S =\Big( \!\frac{L}{1 \, \text{m}}\!\Big) \sqrt{ \left(\!\frac{n+1}{10^2}\!\right)\!\left(\!\frac{\mathcal{F}}{5 \times 10^{5}}\!\right)\!\left(\!\frac{1 \, \mu\text{Hz}}{\Gamma_D}\!\right)}.
\label{eq:def-S}
\end{align}
For the currently achievable value of $n= 10^2$, $h_{\text{noise}}$ can be as low as $6.3 \times 10^{-29} $ which is only 1.7 orders of magnitude above the $\Delta N_{\text{eff}}$ bound on GWs. While the precise trajectory of future developments remains uncertain, it is natural to expect that continued advancements in qumode technology and improvements in microwave cavity quality at cryogenic temperatures could enhance $S$
by 1.7 orders of magnitudes in the near future. It could then push the sensitivity below the $\Delta N_{\text{eff}}$ bound, potentially opening a window into the cosmic gravitational wave background, characterized by a sharp peak at this frequency. Enhancing sensitivity across a broadband spectrum is more challenging and necessitates the use of the qumode configuration given in \cref{eq:Qu-2}. In \cref{fig:detector}, the sensitivity to the narrowband is labeled as \textit{Qu}Grav nb, while the sensitivity to the broadband spectrum appears as \textit{Qu}Grav Qu-bb.


{\bf{Summary and Conclusion.}} In this letter, we propose a novel approach to detecting high-frequency gravitational waves using qumodes, which are used in continuous-variable quantum computation. The detection mechanism relies on the inverse Gertsenshtein effect, wherein a GW is resonantly converted into a single photon within a cavity subjected to a static magnetic field. In this framework, the presence of an $n$-photon qumode leads to an enhancement in the conversion probability by a factor of $n+1$, owing to Bose-Einstein statistics.  The price to pay for the increased sensitivity is the necessity to rebuild the qumode within its coherence time.

Our analysis based on the qumode state defined in \cref{eq:Qu-1} for narrowband spectra shows that, at microwave frequencies and with currently achievable technology, the proposed setup can reach sensitivities up to $6.3 \times 10^{-29}$ \cref{eq:SNR-nb}, i.e., within 1.7 orders of magnitude of the current cosmological bounds. With near-future advancements and improving $L \sqrt{(n+1)\mathcal{F}/\Gamma_D}$, it may even surpass these bounds, opening a window to probe the stochastic GW background of cosmological origin. At optical  regimes the same framework could enhance the sensitivity of existing narrowbound detectors by an order of magnitude (see \cref{eq:h-micro}), potentially allowing them to approach the single-graviton detection threshold. Enhancing sensitivity across a broadband spectrum presents greater challenges and calls for further advancements in the generation and stabilization of qumodes. Using the theoretical benchmark qumode state introduced in \cref{eq:Qu-2}, we compute the corresponding transition rate in \cref{eq:Gamma-bb} and sensitivity in  \cref{eq:SNR-bb}, illustrating its long-term potential for HFGW detection in \cref{fig:detector}.
\vskip 0.1cm

\begin{acknowledgments}
{\bf{Acknowledgments:} } We are grateful to Valerie Domcke, Sebastian Ellis, 
Steven Girvin, and Nicholas Rodd for their valuable comments on our draft, and to Sung-Mook Lee for insightful discussions. The work of A.M. was supported by the Royal Society University Research Fellowship, Grant No. RE22432. The work of D.K. and S.S. was supported by the U.S. Department of Energy, Office of Science, Office of Nuclear Physics, Grants No. DE-FG88ER41450 and DE-SC0012704 and by the U.S. Department of Energy, Office of Science, National Quantum Information Science Research Centers, Co-design Center for Quantum Advantage (C2QA) under Contract No.DE-SC0012704.
\end{acknowledgments}





\clearpage                 
\onecolumngrid             

\setcounter{page}{1}
\renewcommand{\thepage}{S\arabic{page}}

\setcounter{equation}{0}
\renewcommand{\theequation}{S\arabic{equation}}

\setcounter{figure}{0}
\renewcommand{\thefigure}{S\arabic{figure}}

\setcounter{table}{0}
\renewcommand{\thetable}{S\arabic{table}}

\makeatletter
\renewcommand{\theHequation}{S\arabic{equation}}
\renewcommand{\theHfigure}{S\arabic{figure}}
\renewcommand{\theHtable}{S\arabic{table}}
\makeatother

\begin{center}
\textbf{\large Supplemental Material for:\\[3pt]
\emph{QuGrav: Bringing gravitational waves to light with Qumodes}}\\[6pt]
Dmitri E. Kharzeev, Azadeh Maleknejad, and Saba Shalamberidze
\end{center}
\vspace{6pt}





\section{S.1 ~ Noise sources and Bose enhancement}

The purpose of this appendix is to clarify which background processes are, and are not, subject to Bose enhancement. In particular, we wish to distinguish between (i) processes that actually generate photons in the cavity mode and hence can be enhanced by a factor of $(n{+}1)$, and (ii) thermal or detection-related noise sources, which do not scale in this way.  

In quantum optics, the dynamics of a single cavity mode described by the annihilation operator $\hat{b}$, weakly coupled to a thermal bath, can be modeled by a master equation for the cavity density matrix $\rho$ \cite{WallsMilburn2008,Carmichael1999, GardinerZoller2004}
\begin{equation}
\frac{d}{dt}\rho
= \kappa(\bar n_T+1)\,\mathcal{D}[\hat{b}]\,\rho
  + \kappa \bar n_T\,\mathcal{D}[\hat{b}^\dagger]\,\rho,
\qquad
\bar n_T(\omega)=\frac{1}{e^{\hbar\omega/k_B T}-1}.
\end{equation}
Here $\kappa$ is the cavity decay rate, and $\mathcal{D}[o]\rho = o\rho o^\dagger - \tfrac12\{o^\dagger o,\rho\}$ denotes the Lindblad dissipator governing photon loss and gain.  From this master equation, one obtains the evolution of the mean photon number in the cavity,
\begin{equation}
\frac{d}{dt}\langle \hat{b}^\dagger \hat{b}\rangle
= -\kappa\big(\langle \hat{b}^\dagger \hat{b}\rangle - \bar n_T\big),
\label{eq:dn_dt}
\end{equation}
which shows that the cavity occupation relaxes exponentially to the thermal value $\bar n_T$. Importantly, this result holds \emph{independently of the initial state} of the cavity (vacuum, Fock, squeezed, etc.).  This demonstrates that thermal loading of the cavity is not Bose-enhanced by $(n{+}1)$. The steady-state thermal background is determined solely by $\bar n_T$ and $\kappa$, and does not depend on whether the cavity initially contained zero photons, $n$ photons, or any other state.

Having clarified that thermal loading of the cavity is not stimulated, it is natural to ask: which other backgrounds, if any, would actually be subject to the $(n{+}1)$ factor? The answer is that only processes described by an interaction Hamiltonian of the form 
\begin{equation}
H_{\rm int} \;\propto\; \hat{b}^\dagger \,\hat{\mathcal{O}} \;+\; \mathrm{h.c.},
\label{eq:BoseProc}
\end{equation}
where $\hat{b}^\dagger$ creates a photon in the signal mode and $\hat{\mathcal{O}}$ is an operator associated with an external source or another independent system degree of freedom (i.e., not involving 
$\hat{b}^\dagger$ itself), can exhibit Bose stimulation. In such cases, the transition rate is proportional to $(n{+}1)$, since the matrix element involves $\langle n+1|\hat{b}^\dagger|n\rangle = \sqrt{n+1}$.

Beyond the desired graviton--photon conversion, this structure can, in principle, arise from spurious channels such as defect-induced luminescence, nonlinear optical processes, or drive leakage into the cavity. These cases are discussed below.
\begin{enumerate}[i)]
  \item Defect or impurity luminescence: Radiative relaxation of defects into the cavity mode could, in principle, be stimulated, but such processes are strongly suppressed in high-purity superconducting or dielectric cavities, and further reduced by spectral filtering \cite{Haroche2006,Poole2010}.  
  \item Nonlinear optical processes: Channels such as four-wave mixing or parametric down-conversion would also be subject to $(n{+}1)$, but in the absence of strong resonant pumps, their rates are vanishingly small. Using low-nonlinearity materials and avoiding phase-matched drives ensures these remain subleading  \cite{Boyd2008,Kumar1990}.  
  \item Drive leakage or electrical pickup: Leakage into the signal band can seed photons that would then be stimulated. Standard cavity shielding and isolation suppress such leakage below the thermal floor \cite{Pozar2011,Haroche2006}.  
\end{enumerate}
By operating at cryogenic temperatures, using low-loss and low-nonlinearity materials, applying spectral filtering, and eliminating resonant pumps at the signal frequency, these channels can be rendered exponentially suppressed. Under such conditions, the only relevant background is the added noise of the detection process, which does not populate the cavity mode and is therefore not Bose-enhanced.

\section{S.2 ~ n-Photon Fock States: Lifetime and Regeneration}

While \eqref{eq:dn_dt} shows that the mean occupation relaxes to $\bar n_T$ irrespective of the initial state, a prepared Fock state $\ket{n}$ is fragile in a lossy cavity. 
At effectively zero temperature ($\bar n_T\!\approx\!0$), photon loss acts as an amplitude-damping channel. Modeling loss as a beamsplitter of transmissivity $\eta(t)=e^{-\kappa t}$, the survival fidelity of the exact number state obeys \cite{NielsenChuang2000,Haroche2006}
\begin{equation}
F_n(t)=\bra{n}\rho(t)\ket{n}=\eta(t)^{\,n}=e^{-n\kappa t},
\end{equation}
and the mean photon number decays as 
\begin{equation}
\langle \hat{b}^\dagger\hat{b}\rangle(t)=n\,e^{-\kappa t}.
\end{equation}
Hence the effective lifetime of $\ket{n}$ scales as $\tau_n\simeq (n\kappa)^{-1}$: larger $n$ leads to proportionally faster decay. 
At finite temperature, upward and downward jumps generated by $\kappa\bar n_T\,\mathcal D[\hat b^\dagger]$ and $\kappa(\bar n_T+1)\mathcal D[\hat b]$ further shorten this timescale.

To reliably maintain an $n$-photon state for readout, one must regenerate (re-prepare) it on a cycle time shorter than its decay time,
\begin{equation}
\tau_{\rm rep}\ll \tau_n \simeq \frac{1}{n\kappa(\omega)}\quad(\text{and } \tau_{\rm rep}\ll 1/\kappa(\omega) \text{ when } \bar n_T>0),
\end{equation}
with the preparation performed immediately before measurement. The state cannot simply be stored in the cavity; it must be actively refreshed at a rate faster than the intrinsic decay. 

The decay rate depends on frequency through $\kappa(\omega) = \omega/Q(\omega)$, which determines the effective lifetime \cite{Pozar2011} 
\begin{equation}
\tau_n(\omega) \;\simeq\; \frac{Q(\omega)}{n\,\omega}.
\end{equation}
Thus, even if the quality factor $Q(\omega)$ were independent of frequency, the lifetime $\tau_n(\omega)$ would still fall off inversely with $\omega$.
 At microwave and optical frequencies, very large $Q$ values can suppress $\kappa(\omega)$ sufficiently that regeneration is experimentally feasible. In contrast, at x-ray and higher frequencies the linear growth of $\kappa(\omega)$ with $\omega$ makes $\tau_n$ extremely short, and achieving $\tau_{\rm rep}\ll\tau_n$ becomes essentially impossible, independent of resonator technology.

We end this appendix with a brief summary of different strategies that have been developed to regenerate (re-prepare) $n$-photon Fock states before readout: 

\begin{enumerate}[i)]
  \item Heralded spontaneous parametric down-conversion (SPDC):  
  Heralded detection of one photon projects the partner mode into a Fock state. Extensions using multiplexing allow approximate preparation of small-$n$ states \cite{HongMandel1986,Lvovsky2009}.

  \item Photon addition and subtraction:  
  Conditional operations based on nonlinear interactions allow photon creation ($\hat b^\dagger$) or annihilation ($\hat b$) on demand, enabling the stepwise preparation of higher-number states \cite{Zavatta2004,Ourjoumtsev2007}.

  \item Qubit--cavity swaps in circuit QED (microwave domain):  
  Strong coupling between a superconducting qubit and a cavity mode allows deterministic loading of photons. By repeating qubit excitations and swaps, Fock states up to $n\!\sim\!15$ have been demonstrated \cite{Hofheinz2008,Hofheinz2009}.

  \item Quantum dots and atom--cavity emission:  
  Single emitters such as quantum dots or trapped atoms can be driven to release photons sequentially into the same mode \cite{Santori2002,McKeever2004}.

  \item Multiplexed probabilistic sources:  
  Combining many heralded sources with active feed-forward (in the spirit of boson sampling) allows the generation of higher-$n$ states, though at significant resource cost \cite{Maunz2007,GimenoSegovia2015}.
\end{enumerate}

\section{S.3 ~ Coherent states vs.\ Qumodes}

To clarify the practical advantage of Qumodes over conventional coherent states, we compare their respective signal-to-noise performance within the same intracavity energy budget.  
A coherent state with mean photon number $n$ enhances the generation rate of signal photons by the Bose factor $(n{+}1)$. However, its detectability is limited by the intrinsic measurement noise of the detection process. Denoting by $n_{\rm add}$ the total added quanta referred to the cavity output (shot noise, amplifier noise, and technical noise lumped together), the best achievable sensitivity with a coherent state at fixed intracavity energy is \cite{WallsMilburn2008}
\begin{equation}
\mathrm{SNR}_{\rm coh} \ \propto\ \frac{\Delta N_{\rm sig}}{\sqrt{n+n_{\rm add}}}
\ \simeq\ \frac{(n+1)}{\sqrt{n+n_{\rm add}}}
\ \xrightarrow[]{\,n\gg n_{\rm add}\,}\ \sqrt{n}\,,
\label{eq:SNRcoh}
\end{equation}
which is the well-known standard quantum limit (SQL). In addition, a bright classical carrier populates the signal band and imports amplitude/phase noise, leakage, and other systematics that scale with the drive power; increasing $n$ further risks heating (quasiparticles, dielectric loss) and spurious mixing.  

By contrast, nonclassical Qumodes preserve a nearly \emph{dark channel}: the signal rate is still enhanced by $(n{+}1)$, but the read-out noise remains unchanged. A detailed discussion of the Qumode construction and its properties can be found in Appendix~S1 and S2.  At fixed stored energy, this yields
\begin{equation}
\mathrm{SNR}_{\rm Qumode} \ \propto\ (n+1)\,,
\label{eq:SNRQumode}
\end{equation}
i.e., \ linear scaling with $n$, surpassing the SQL and providing robust background discrimination.  In summary, while both coherent states and Qumodes amplify the \emph{rate} via the same Bose factor $(n{+}1)$, only Qumodes decouple the signal enhancement from added noise. This allows higher SNR, sub-SQL performance, and cleaner identification of extremely weak processes such as high-frequency graviton--photon conversion.


\begin{thebibliography}{99}

\bibitem{Gertsenshtein}
M. E. Gertsenshtein,
``Wave Resonance of Light and Gravitational Waves,''
Sov. Phys. JETP 14 (1962) 84

\bibitem{Ejlli:2019bqj}
A.~Ejlli, D.~Ejlli, A.~M.~Cruise, G.~Pisano and H.~Grote,
``Upper limits on the amplitude of ultra-high-frequency gravitational waves from graviton to photon conversion,''
Eur. Phys. J. C \textbf{79}, no.12, 1032 (2019)


\bibitem{OSQAR:2015qdv}
R.~Ballou \textit{et al.} [OSQAR],
``New exclusion limits on scalar and pseudoscalar axionlike particles from light shining through a wall,''
Phys. Rev. D \textbf{92}, no.9, 092002 (2015)

\bibitem{Albrecht:2020ntd}
C.~Albrecht, S.~Barbanotti, H.~Hintz, K.~Jensch, R.~Klos, W.~Maschmann, O.~Sawlanski, M.~Stolper and D.~Trines,
``Straightening of Superconducting HERA Dipoles for the Any-Light-Particle-Search Experiment ALPS II,''
EPJ Tech. Instrum. \textbf{8}, no.1, 5 (2021)


\bibitem{CAST:2017uph}
V.~Anastassopoulos \textit{et al.} [CAST],
``New CAST Limit on the Axion-Photon Interaction,''
Nature Phys. \textbf{13}, 584-590 (2017)

\bibitem{IAXO:2020wwp}
A.~Abeln \textit{et al.} [IAXO],
``Conceptual design of BabyIAXO, the intermediate stage towards the International Axion Observatory,''
JHEP \textbf{05}, 137 (2021)

\bibitem{IAXO:2019mpb}
E.~Armengaud \textit{et al.} [IAXO],
``Physics potential of the International Axion Observatory (IAXO),''
JCAP \textbf{06}, 047 (2019)

\bibitem{Beacham:2019nyx}
J.~Beacham, C.~Burrage, D.~Curtin, A.~De Roeck, J.~Evans, J.~L.~Feng, C.~Gatto, S.~Gninenko, A.~Hartin and I.~Irastorza, \textit{et al.}
``Physics Beyond Colliders at CERN: Beyond the Standard Model Working Group Report,''
J. Phys. G \textbf{47}, no.1, 010501 (2020)


\bibitem{Aggarwal:2025noe}
N.~Aggarwal, O.~D.~Aguiar, D.~Blas, A.~Bauswein, G.~Cella, S.~Clesse, A.~M.~Cruise, V.~Domcke, S.~Ellis and D.~G.~Figueroa, \textit{et al.}
``Challenges and Opportunities of Gravitational Wave Searches above 10 kHz,''
[arXiv:2501.11723 [gr-qc]].

\bibitem{Blais:2020wjs}
A.~Blais, A.~L.~Grimsmo, S.~M.~Girvin and A.~Wallraff,
``Circuit quantum electrodynamics,''
Rev. Mod. Phys. \textbf{93}, no.2, 025005 (2021)

\bibitem{Stavenger:2022wzz}
T.~J.~Stavenger, E.~Crane, K.~C.~Smith, C.~T.~Kang, S.~M.~Girvin and N.~Wiebe,
``C2QA - Bosonic Qiskit,''
[arXiv:2209.11153 [quant-ph]].


\bibitem{Araz:2024kkg}
J.~Y.~Araz, M.~Grau, J.~Montgomery and F.~Ringer,
``Hybrid quantum simulations with qubits and qumodes on trapped-ion platforms,''
Phys. Rev. A \textbf{112}, no.1, 012620 (2025)

\bibitem{miesner1998bosonic}
Miesner, H. J., Stamper-Kurn, D. M., Andrews, M. R., Durfee, D. S., Inouye, S., Ketterle, W.  
``Bosonic stimulation in the formation of a Bose-Einstein condensate,''
Science, 279(5353), 1005-1007, (1998).

\bibitem{lu2023bosonic}
Lu, Yu-Kun, Yair Margalit, and Wolfgang Ketterle, 
``Bosonic stimulation of atom–light scattering in an ultracold gas,''
Nature Physics 19.2 (2023): 210-214.

\bibitem{qu-exp1}
R.~W.~Heeres, B.~Vlastakis, E.~Holland, S.~Krastanov, V.~V.~Albert, L.~Frunzio, L.~Jiang and R.~J.~Schoelkopf,
``Cavity State Manipulation Using Photon-Number Selective Phase Gates,''
Phys. Rev. Lett. \textbf{115}, no.13, 137002 (2015)

\bibitem{qu-exp2}
A.~Eickbusch, V.~Sivak, A.~Z.~Ding, S.~S.~Elder, S.~R.~Jha, J.~Venkatraman, B.~Royer, S.~M.~Girvin, R.~J.~Schoelkopf and M.~H.~Devoret,
``Fast universal control of an oscillator with weak dispersive coupling to a qubit,''
Nature Phys. \textbf{18}, no.12, 1464-1469 (2022)

\bibitem{Cheng:2022wgy}
R.~Cheng, Y.~Zhou, S.~Wang, M.~Shen, T.~Taher and H.~X.~Tang,
``A 100-pixel photon-number-resolving detector unveiling photon statistics,''
Nature Photon. \textbf{17}, no.1, 112-119 (2023)

\bibitem{Ratzinger:2024spd}
W.~Ratzinger, S.~Schenk and P.~Schwaller,
``A coordinate-independent formalism for detecting high-frequency gravitational waves,''
JHEP \textbf{08}, 195 (2024)

\bibitem{Tobar:2023ksi}
G.~Tobar, S.~K.~Manikandan, T.~Beitel and I.~Pikovski,
``Detecting single gravitons with quantum sensing,''
Nature Commun. \textbf{15}, no.1, 7229 (2024)


\bibitem{Agrawal:2023umy}
A.~Agrawal, A.~V.~Dixit, T.~Roy, S.~Chakram, K.~He, R.~K.~Naik, D.~I.~Schuster and A.~Chou,
``Stimulated Emission of Signal Photons from Dark Matter Waves,''
Phys. Rev. Lett. \textbf{132}, no.14, 140801 (2024)

\bibitem{Dyson:2013hbl}
F.~Dyson,
``Is a graviton detectable?,''
Int. J. Mod. Phys. A \textbf{28}, 1330041 (2013)


\bibitem{Rothman:2006fp}
T.~Rothman and S.~Boughn,
``Can gravitons be detected?,''
Found. Phys. \textbf{36}, 1801-1825 (2006)


\bibitem{Carney:2023nzz}
D.~Carney, V.~Domcke and N.~L.~Rodd,
``Graviton detection and the quantization of gravity,''
Phys. Rev. D \textbf{109}, no.4, 044009 (2024)

\bibitem{Schuster:2007fwf}
D.~I.~Schuster, A.~A.~Houck, J.~A.~Schreier, A.~Wallraff, J.~M.~Gambetta, A.~Blais, L.~Frunzio, J.~Majer, B.~Johnson and M.~H.~Devoret, \textit{et al.}
``Resolving photon number states in a superconducting circuit,''
Nature \textbf{445}, no.7127, 515-518 (2007)

\bibitem{Reagor:2013}
M.~Reagor, H.~Paik, G.~Catelani, L.~Sun, C.~Axline, E.~Holland, I.~M.~Pop, N.~A.~Masluk, T.~Brecht and L.~Frunzio, \textit{et al.}
``Reaching 10{\,}ms single photon lifetimes for superconducting aluminum cavities,''
Appl. Phys. Lett. \textbf{102}, no.19, 192604 (2013)
doi:10.1063/1.4807015

\bibitem{Kessler}
T.~Kessler, C.~Hagemann, C.~Grebing, T.~Legero, U.~Sterr, F.~Riehle,
M.~J.~Martin, L.~Chen and J.~Ye,
``A sub-40-mHz-linewidth laser based on a silicon single-crystal optical cavity,''
\emph{Nature Photonics} \textbf{6}, no.~10, 687--692 (2012)


\bibitem{Lescanne:2020awk}
R.~Lescanne, S.~Del\'eglise, E.~Albertinale, U.~R\'eglade, T.~Capelle, E.~Ivanov, T.~Jacqmin, Z.~Leghtas and E.~Flurin,
``Irreversible Qubit-Photon Coupling for the Detection of Itinerant Microwave Photons,''
Phys. Rev. X \textbf{10}, no.2, 021038 (2020)

\bibitem{Dixit:2020ymh}
A.~V.~Dixit, S.~Chakram, K.~He, A.~Agrawal, R.~K.~Naik, D.~I.~Schuster and A.~Chou,
``Searching for Dark Matter with a Superconducting Qubit,''
Phys. Rev. Lett. \textbf{126}, no.14, 141302 (2021)

\bibitem{Ringwald:2020ist}
A.~Ringwald, J.~Sch\"utte-Engel and C.~Tamarit,
``Gravitational Waves as a Big Bang Thermometer,''
JCAP \textbf{03}, 054 (2021)


\bibitem{SM}
See Supplemental Material for details of
the extended discussion, which includes
Refs. \cite{WallsMilburn2008-}- \cite{GimenoSegovia2015-}.

\bibitem{WallsMilburn2008-} D. F. Walls and G. J. Milburn, 
\textit{Quantum Optics}, 2nd ed. (Springer, Berlin, 2008).

\bibitem{Carmichael1999} H. J. Carmichael, 
\textit{Statistical Methods in Quantum Optics 1} 
(Springer, Berlin, 1999).

\bibitem{GardinerZoller2004}
C.~W. Gardiner and P.~Zoller, 
\textit{Quantum Noise: A Handbook of Markovian and Non-Markovian Quantum Stochastic Methods with Applications to Quantum Optics}, 
3rd ed.\ (Springer, Berlin, Heidelberg, 2004).

\bibitem{Haroche2006} S.~Haroche and J.-M.~Raimond, 
\textit{Exploring the Quantum: Atoms, Cavities, and Photons} 
(Oxford University Press, 2006).

\bibitem{Poole2010} C.~P.~Poole  \textit{et al.},
\textit{Superconductivity}, 2nd ed.
(Academic Press, 2010).

\bibitem{Boyd2008} R.~W.~Boyd, 
\textit{Nonlinear Optics}, 3rd ed. (Academic Press, 2008).



\bibitem{Kumar1990} P.~Kumar, 
“Quantum frequency conversion,” 
Opt. Lett. \textbf{15}, 1476 (1990).

\bibitem{Pozar2011} D.~M.~Pozar, 
\textit{Microwave Engineering}, 4th ed. (Wiley, 2011).

\bibitem{NielsenChuang2000} 
M.~A. Nielsen and I.~L. Chuang, 
\textit{Quantum Computation and Quantum Information}, 
(Cambridge University Press, 2000).



\bibitem{HongMandel1986} C.~K. Hong and L. Mandel, 
“Experimental realization of a localized one-photon state,” 
Phys. Rev. Lett. \textbf{56}, 58 (1986).

\bibitem{Lvovsky2009} A.~I. Lvovsky and M.~G. Raymer, 
“Continuous-variable optical quantum-state tomography,” 
Rev. Mod. Phys. \textbf{81}, 299 (2009).



\bibitem{Zavatta2004} A. Zavatta, S. Viciani, and M. Bellini, 
“Quantum-to-classical transition with single-photon-added coherent states of light,” 
Science \textbf{306}, 660 (2004).

\bibitem{Ourjoumtsev2007} A. Ourjoumtsev \textit{et al.}, 
“Generation of optical ‘Schrödinger cats’ from photon number states,” 
Nature \textbf{448}, 784 (2007).

\bibitem{Hofheinz2008} M. Hofheinz \textit{et al.}, 
“Generation of Fock states in a superconducting quantum circuit,” 
Nature \textbf{454}, 310 (2008).



\bibitem{Hofheinz2009} M. Hofheinz \textit{et al.}, 
“Synthesizing arbitrary quantum states in a superconducting resonator,” 
Nature \textbf{459}, 546 (2009).



\bibitem{Santori2002} C. Santori \textit{et al.}, 
“Indistinguishable photons from a single-photon device,” 
Nature \textbf{419}, 594 (2002).

\bibitem{McKeever2004} J. McKeever \textit{et al.}, 
“Deterministic generation of single photons from one atom trapped in a cavity,” 
Science \textbf{303}, 1992 (2004).



\bibitem{Maunz2007} P. Maunz \textit{et al.}, 
“Quantum interference of photon pairs from two remote trapped atomic ions,” 
Nature Phys. \textbf{3}, 538 (2007).

\bibitem{GimenoSegovia2015-} M. Gimeno-Segovia \textit{et al.}, 
“From three-photon Greenberger–Horne–Zeilinger states to ballistic universal quantum computation,” 
Phys. Rev. Lett. \textbf{115}, 020502 (2015).


\end{thebibliography}

\begin{thebibliography}{99}

\bibitem{WallsMilburn2008} D. F. Walls and G. J. Milburn, 
\textit{Quantum Optics}, 2nd ed. (Springer, Berlin, 2008).

\bibitem{Carmichael1999} H. J. Carmichael, 
\textit{Statistical Methods in Quantum Optics 1} 
(Springer, Berlin, 1999).

\bibitem{GardinerZoller2004}
C.~W. Gardiner and P.~Zoller, 
\textit{Quantum Noise: A Handbook of Markovian and Non-Markovian Quantum Stochastic Methods with Applications to Quantum Optics}, 
3rd ed.\ (Springer, Berlin, Heidelberg, 2004).

\bibitem{Haroche2006} S.~Haroche and J.-M.~Raimond, 
\textit{Exploring the Quantum: Atoms, Cavities, and Photons} 
(Oxford University Press, 2006).

\bibitem{Poole2010} C.~P.~Poole  \textit{et al.},
\textit{Superconductivity}, 2nd ed.
(Academic Press, 2010).

\bibitem{Boyd2008} R.~W.~Boyd, 
\textit{Nonlinear Optics}, 3rd ed. (Academic Press, 2008).

\bibitem{Kumar1990} P.~Kumar, 
“Quantum frequency conversion,” 
Opt. Lett. \textbf{15}, 1476 (1990).

\bibitem{Pozar2011} D.~M.~Pozar, 
\textit{Microwave Engineering}, 4th ed. (Wiley, 2011).

\bibitem{NielsenChuang2000} 
M.~A. Nielsen and I.~L. Chuang, 
\textit{Quantum Computation and Quantum Information}, 
(Cambridge University Press, 2000).

\bibitem{HongMandel1986} C.~K. Hong and L. Mandel, 
“Experimental realization of a localized one-photon state,” 
Phys. Rev. Lett. \textbf{56}, 58 (1986).

\bibitem{Lvovsky2009} A.~I. Lvovsky and M.~G. Raymer, 
“Continuous-variable optical quantum-state tomography,” 
Rev. Mod. Phys. \textbf{81}, 299 (2009).

\bibitem{Zavatta2004} A. Zavatta, S. Viciani, and M. Bellini, 
“Quantum-to-classical transition with single-photon-added coherent states of light,” 
Science \textbf{306}, 660 (2004).

\bibitem{Ourjoumtsev2007} A. Ourjoumtsev \textit{et al.}, 
“Generation of optical ‘Schrödinger cats’ from photon number states,” 
Nature \textbf{448}, 784 (2007).

\bibitem{Hofheinz2008} M. Hofheinz \textit{et al.}, 
“Generation of Fock states in a superconducting quantum circuit,” 
Nature \textbf{454}, 310 (2008).

\bibitem{Hofheinz2009} M. Hofheinz \textit{et al.}, 
“Synthesizing arbitrary quantum states in a superconducting resonator,” 
Nature \textbf{459}, 546 (2009).

\bibitem{Santori2002} C. Santori \textit{et al.}, 
“Indistinguishable photons from a single-photon device,” 
Nature \textbf{419}, 594 (2002).

\bibitem{McKeever2004} J. McKeever \textit{et al.}, 
“Deterministic generation of single photons from one atom trapped in a cavity,” 
Science \textbf{303}, 1992 (2004).

\bibitem{Maunz2007} P. Maunz \textit{et al.}, 
“Quantum interference of photon pairs from two remote trapped atomic ions,” 
Nature Phys. \textbf{3}, 538 (2007).

\bibitem{GimenoSegovia2015} M. Gimeno-Segovia \textit{et al.}, 
“From three-photon Greenberger–Horne–Zeilinger states to ballistic universal quantum computation,” 
Phys. Rev. Lett. \textbf{115}, 020502 (2015).


\end{thebibliography}
\end{document}